\documentclass[aps,prl,twocolumn,unsortedaddress,superscriptaddress]{revtex4-1}
\usepackage{ucs}
\usepackage[utf8]{inputenc}
\usepackage{amsfonts}
\usepackage{amsmath}
\usepackage[american]{babel}
\usepackage[T1]{fontenc}
\usepackage[dvips]{graphicx}
\usepackage{epsfig}
\date{\today}

\begin{document}
\author{Xavier Illa}
\affiliation{Departament d’Estructura i Constituents de la Mat\`eria. Facultat
de F\'isica.  Universitat de Barcelona.  Diagonal, 647, E-08028 Barcelona, Catalonia,
Spain}
\affiliation{Aalto University, School of Science, Department of Applied
 Physics, P.O.  Box 14100, FI-00076 AALTO, Finland}
\author{Antti Puisto}
\affiliation{Aalto University, School of Science, Department of Applied
 Physics, P.O.  Box 14100, FI-00076 AALTO, Finland}
\author{Arttu Lehtinen}
\affiliation{Aalto University, School of Science, Department of Applied
 Physics, P.O.  Box 14100, FI-00076 AALTO, Finland}
\author{Mikael Mohtaschemi}
\affiliation{Aalto University, School of Science, Department of Applied
 Physics, P.O.  Box 14100, FI-00076 AALTO, Finland}
\author{Mikko J. Alava}
\affiliation{Aalto University, School of Science, Department of Applied
 Physics, P.O.  Box 14100, FI-00076 AALTO, Finland}

\title{Transient Shear Banding in Time-dependent Fluids}
\begin{abstract}
We study the dynamics of shear-band formation and evolution using a simple
rheological model. The description couples the local structure and
viscosity to the applied shear stress. We consider in detail the
Couette geometry, where the model is solved iteratively with the
Navier-Stokes equation to obtain the time-evolution of the local
velocity and viscosity fields.  It is found that the underlying
reason for dynamic effects is the non-homogeneous shear
distribution, which is amplified due to a positive feedback between
the flow field and the viscosity response of the shear thinning
fluid. This offers a simple explanation for the recent observations
of Transient Shear Banding in time-dependent fluids. Extensions to
more complicated rheological systems are considered.
\end{abstract}

\maketitle

The property that characterizes complex fluids is their non-trivial rheology,
shear rate -- stress relation.  They are generally further categorized into
shear thinning or shear thickening fluids.  Both cases are additionally
complicated by time dependence.  Due to the stress-shear interaction, already
small perturbations in the local stress can result in a positive feedback with
the flow promoting shear instabilities in each case
\cite{Head2002,Fielding2004}.  The understanding of complex fluids is of
enormous importance for many practical applications \cite{Braun1999} and the
theory touches on many branches of physics.  Recent advances allow to follow
the suspension local velocity during a standard rheological experiment
\cite{coussot, manneville}.  Quantifying the local flow field simultaneously
with rheological measurements gives the possibility to measure both the
intrinsic and apparent rheology.  This has lead to the discovery that a
heterogeneous shear distribution in samples during such tests is ubiquitous. 
Shear banding~\cite{ovarlez} has been observed in many systems composed of
substantially different building blocks, such as colloidal glasses, wormlike
micelles, foams and granular matter~\cite{schall_2010}.  The current viewpoint,
both phenomenologically and theoretically, is that a non-monotonic intrinsic
flow curve, is what is common to most of these materials~\cite{ovarlez,
Porte1997}, but also other mechanisms have been suggested~\cite{besseling}.

A branch of complex fluids are the simple yield stress fluids~\cite{ragouilliaux}. 
These materials do not show aging phenomena (thixotropy).  Therefore they
are expected to have a monotonic intrinsic flow curve, and a steady state
without shear bands \cite{Fall2010}.  However, recent
experiments~\cite{divoux} display shear banding during startup flows in a
rotational rheometer indicating time-dependent behavior.  These, so called
transient shear bands, can be very long lasting, but eventually vanish with
a homogeneous steady state.  The transient shear banding phenomenon tests
our fundamental understanding of non-Newtonian fluids, and is also important
for industrial processes and simply for understanding usual rheological
measurements.  A particular feature of the transient shear banding is that
it appears to exhibit scaling familiar from critical phenomena: the time it
takes for the transient to disappear (fluidization time $\tau_f$) is a
power-law function of the shear rate or applied stress \cite{divoux}.

For the reasons leading to transient shear banding, see the summary by Adams
{\it et al.} \cite{fielding}, three main candidates are offered: i) with
certain parameters, the dynamical equations are unstable amplifying
small perturbations, which slowly quench towards the homogeneous steady state
\cite{fielding2003}.  ii) The fluid flow curve is time dependent, and can, at
different times, have non-monotonic shape \cite{hayes2010}.  iii) Elastic
stress overshoots cause instability in the flow \cite{marrucci}.  Some
theoretical models appear to produce transient shear banding including shear
transformation zone theories \cite{manning}, a modified soft glassy rheology
model \cite{moorcroft}, a simplified fluidity model \cite{moorcroft}, and a
mesoscopic model of plasticity \cite{jagla}.  Such models of transient shear banding share the
property of time-dependent reduction of the local stress under shear
\cite{manning,moorcroft,jagla} as explained in Ref.~\cite{Moorcroft_2012}. 
Literature reports experimental evidence of time-dependent rheology in carbopol
gels \cite{divoux} especially at small shear rates, as well as in other simple
yield stress fluids \cite{moller_2009}, appearing as slight hysteresis in the
flow curves.  Further details of the time-dependent flow curve hysteresis
related to carbopol gels is reported in Ref.~\cite{divoux_sm2011}.  Stronger
hysteresis is inherent to thixotropic fluids \cite{moller_2009}.  Indeed,
transient shear banding has been recently found also there well above the
critical shear rates \cite{martin} indicating that transient shear banding
could be a general property of soft glassy materials.

Motivated by these findings, we consider here a structural model to find the
main ingredients of transient shear banding.  It is for a simple time-dependent
Newtonian fluid, on purpose neglecting other complications present in yield
stress fluids and thixotropic fluids, such as elastic and yield stresses and
critical shear rates.  Spatial resolution, necessary for the transient shear
banding, is obtained when the rheological model is coupled to the Navier-Stokes
equation for laminar flow in a concentric cylinder Couette device.  The
transient shear banding here occurs due to the shear rate - viscosity coupling
in the Navier-Stokes solution.  We will demonstrate that such transients
initiate since the shear stress exhibits a finite gradient in any rotational
geometry.  This gradient is amplified by the interaction with the shear
thinning fluid.  In the model, the timescales associated with the relaxation of
the structure depend on the shear rate.  A higher shear rate implies faster
dynamics creating, with the coupled Navier-Stokes equations, an amplifying
effect during the startup flow.

{\it Model.} --- Most microscopic descriptions of the structure of a complex
fluid are purely phenomenological, and based on a kinetic relaxation equation. 
The simplest $\lambda$-models stipulate an evolution equation for a structural
parameter, $\lambda$ \cite{coussot, mewis}.  This describes the internal order,
such as the state of aggregation in colloids or the alignment of particles with
shear \cite{coussot}.  Usually, shear and/or temperature influences the
temporal evolution of $\lambda$ \cite{mewis}.  The macroscopic rheology is
obtained coupling $\lambda$ to a constitutive equation.  Variants of
$\lambda$-models can be used to describe the flow curves of thixotropic, simple
yield stress fluids, and shear thinning fluids \cite{moller,coussot}.  They
generally assume homogeneous flow, and thus are not applicable to shear bands.

In aggregating suspensions and microgels a portion of the liquid is trapped due
to the presence of solid structures~\cite{heath,piau2007}.  This is infact also
the interpretation shared in the literature for carbopol microstructure, which
is observed to be formed of elastic sponge-like
elements~\cite{piau2007,oppong2006}.  Therefore, instead of using an abstract
structure parameter (with no direct physical interpretation) it is more
sensible to select the immobilized volume fraction $\phi$, which describes the
jammed fluid, as it has been done in approaches using the Population Balance 
Equations \cite{heath}.  The simplest time-dependent rheology follows by including a
temporal relaxation of the volume fraction to a steady state.  Unlike the
relaxation rate, the steady state volume fraction is independent of the applied
shear rate corresponding to a Newtonian fluid.  Such a model can be thought as
a Taylor-expansion of the dynamical equation for the volume around the
steady-state.  
The resulting kinetic equation is given 
\begin{equation}
\frac{d\phi}{dt} = \frac{A_b}{(\mu/\mu_o)^{m}} + \left( A_s - B_s \phi \right) \left(\frac{\dot\gamma}{\dot
\gamma_0}\right)^{k}, \label{eq:simple}
\end{equation}
where $A_s$ ($B_s$) is the kinetic constant for the shear growth
(destruction), $\dot\gamma$ is the magnitude of the shear rate, $k$ and $\dot\gamma_0$ (set to unity) both relate to the volume
fraction sensitivity to shearing. 
$A_b$, $\mu_o$, and $m$ describe the growth of jammed volume
fraction due to the shear independent motion of the structure elements. The special case of
$m = k$, presents a simple yield stress fluid,
$m > k$ gives a non-monotonic flow curve indicating thixotropy, and
$m < k$ produces shear thinning behavior. In what follows, we fix the parameters $A_s$
and $B_s$ to 0.665 and 1.0, respectively.
Since we are concentrating on a minimum model showing transient shear banding, in the following we
set $A_b = 0$, and parametrize the initial volume fraction instead of
specifying the sample history and the associated parameters. It is well known that in any practical
experiment the initial state depends on the sample history, and the shear independent structure dynamics.
Therefore special attention has to be payed to the measuring protocol.
The shear independent terms dominate the structure evolution at small shear rates making the flow
curve non-Newtonian, but have negligible influence in the transient shear
banding regime. There, it is reasonable to assume that both, the structure
growth and destruction terms are determined by the shear rate justifying such approximation.  The exact
form of the rate kernels are presently unclear and here we choose a simple power-law
dependence.  The scaling factors $\dot\gamma_0$ and $\mu_o$ make the equation unitless, and
could be incorporated to the kernels $A_s$, $A_b$, and $B_s$ equally well.

We use the well-known Krieger-Dougherty constitutive
equation~\cite{krieger}, that reads 
\begin{equation} 
\mu(\phi) = \mu_0\left(1-\frac{\phi}{\phi_m}\right)^{-\eta}, \label{eq:visc}
\end{equation}
where $\phi_m$ is the jamming volume fraction, $\mu_0$ the liquid viscosity,
and $\eta$ gives the scaling of the viscosity.  From now on these quantities
are set to $\phi_m = 0.68$ (random sphere packing), $\mu_0 = 1$ mPas (water),
and $\eta = 1.82$ \cite{quemada}.  These values relate to the initial
conditions at which the transient shear banding appears, but are irrelevant
e.g.  for the scaling relations of the fluidization times.

\begin{figure}[htb]
\begin{center}
  \epsfig{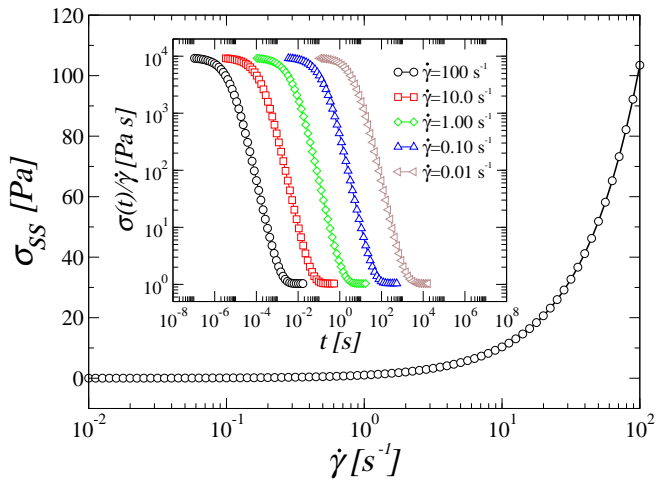}
\end{center}
 \caption{(Color online) The steady state flow curve, showing Newtonian response, with $A_s =
 0.665$, $B_s = 1.000$.  Inset: the transient behavior of the viscosity. $\phi_0 
= 0.6799$, $k=1.5$.
\label{FIG1}}
\end{figure}

The steady state volume fraction is $\phi_{ss} = \frac{A_s}{B_s}$.
As a consequence of leaving the shear independent terms out of the kinetic
equation, the resulting steady state volume fraction is independent of the
shear rate.  As mentioned before, these shear independent processes are the ones that drive the
system out of steady state implied by Eq. (1) when not sheared.  This effect
is incorporated to the model by simply initializing the system to the desired
volume fraction before starting the structure evolution.  
For shear thinning fluids having no yield stress, the present approximation is valid
for all shear rates. Fluids having a yield stress, are properly described in the range
where shear localization does not appear i. e. the fluidization is complete.
%
Besides fixing the steady state, $A_s$ and $B_s$ determine the rate of relaxation. 
Plugging $\phi_{ss}$ to the constitutive equation and applying the Newtonian
assumption $\sigma_{ss} = \mu(\phi_{ss}) \dot\gamma$, gives the Newtonian
steady state flow curve in Fig.~\ref{FIG1}.  The stress shows time and shear
rate dependent exponential relaxation (the Inset), when started from $\phi_{o}
\neq \phi_{ss}$, since the structure evolution follows Eq.  (\ref{eq:simple}).

For the spatial resolution we couple the model with a continuum description of
the fluid flow in a Couette rheometer geometry~\cite{roussel2004}.  The
incompressible Navier-Stokes equation for laminar flow in that has the
analytical one-dimensional (radial) solution \cite{philips} %
\begin{equation}
 \dot \gamma(r) = \frac{\Omega_b-\Omega_a}{\left[\int_{R_a}^{R_b} \frac{1}{r^3
\mu(r)}dr\right]} \cdot \frac{1}{r^2 \mu(r)},
\label{eq:NS}
\end{equation}
where $r$ is the radial distance from the cell center, $\dot\gamma(r)$ is the
local shear rate, $\Omega_a~(\Omega_b)$ is the angular velocity of the inner
(outer) cylinder, and $R_a~(R_b)$ is the radius of the inner (outer) cylinder. 
Eq.~(\ref{eq:NS}) implies that only the relative angular velocity of the two
cylinders matters.  In order to mimic the experiments of Ref.~\cite{divoux}
only the inner cylinder is rotated and the radii are set to $R_a = 23.9$ mm and
$R_b = 25$ mm.  The Eq.~(\ref{eq:NS}) implies a spatial dependence for the
volume fraction $\phi(t,r)$.  To this end, a radial discretization is applied
with a uniform grid with N sampling points, at each of which a separate $\phi$
is evolved.  Thus, the time evolution of the local flow field is obtained when
the Equations ~(\ref{eq:simple}) and ~(\ref{eq:NS}) are iteratively solved
using a forward Euler algorithm under a constant global shear rate. 
Solving the structure and flow field evolutions this way assumes that the time-scales of
the inertial effects and the structure kinetics is very different.  The flow
field quickly adapts the changes caused by the slowly varying time-dependent
viscosity.

The global shear rate is defined as the radial average shear rate
$\int_{R_a}^{R_b} \dot\gamma(r)/(R_b-R_a)$ for a Newtonian fluid in the Couette
geometry.  This reads
\begin{equation}
 \langle \dot\gamma \rangle  = \left(\Omega_b - \Omega_a \right)\frac{ 2 R_a R_b}{R_b^2-R_a^2}.
\label{avg_sr}
\end{equation}
Other definitions for the ``engineering'' shear rate are also used
\cite{estelle2008,becu2004}.  All such are linearly proportional to the
difference of the angular velocities, $\Omega_b - \Omega_a$.

\begin{figure}[htb]
\begin{center}
  \epsfig{file=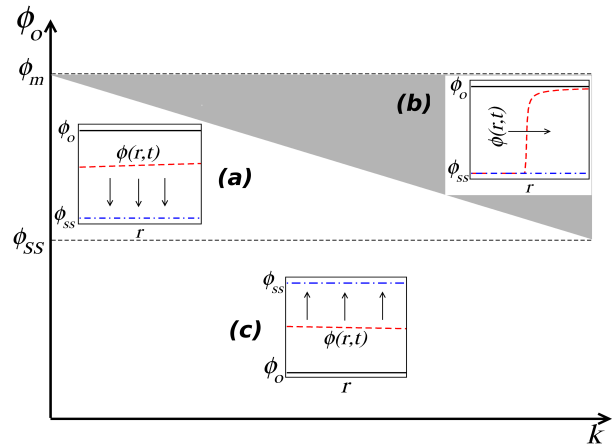,width=8.0cm,clip=}
\end{center}
\caption{\label{FIG2}  (Color online) A schematic phase diagram of the
instability/stability regimes as a function of $k$ and $\phi_0$.  In regions
(a) and (c) spatially homogeneous relaxation is observed, whereas region (b)
exhibits transient shear banding.  }
\end{figure}

{\it Results}. --- Varying the initial volume fraction and the kinetic exponent
$k$ uncovers three different startup flow scenarios as illustrated in the
schematic phase diagram shown in Fig.~\ref{FIG2}.  An almost homogeneous
relaxation of the volume fraction profile arises at initial volume fractions
above the steady state value and low kinetic exponents (Fig.~\ref{FIG2}(a)). 
Increasing either makes the transient shear bands to appear
(Fig.~\ref{FIG2}(b)).  At initial volume fractions below the steady state,
there is no transient shear bands (Fig.~\ref{FIG2}(c)).  The boundaries of that phase (grey area)
are qualitative, as the transition from the homogeneous relaxation
Fig.~\ref{FIG2}(a) to the transient shear banding Fig.~\ref{FIG2}(b) occurs
smoothly.

\begin{figure}[htb]
\begin{center}
  \epsfig{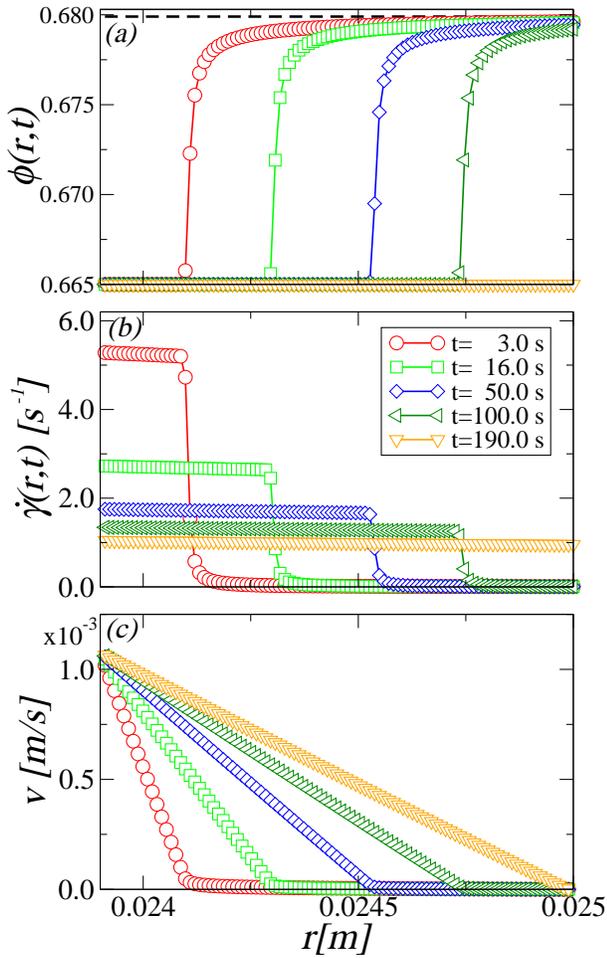}
\end{center}
\caption{\label{FIG3}  (Color online) The volume fraction (upper panel) and
corresponding shear rate (lower panel) during the startup flow.  $A_s = 0.665$,
$B_s = 1.000$, $k = 1.5$, $\langle \dot\gamma \rangle = 1.0$ s$^{-1}$ and
$\phi(r,t=0) = 0.6799$.}
\end{figure}

For a given value of $k$, the initial value $\phi_o$ is chosen in order to be
in the transient shear banding region (grey area in Fig.~\ref{FIG2}).  Thus, the simulation starts
close to jamming, at $\phi_o = \phi(r,t=0) \sim \phi_m$.  Fig.~\ref{FIG3}(b)
shows the local shear rates corresponding to the evolution of the volume
fraction plotted in Fig.~\ref{FIG3}(a), from the numerical solution of
Eqs.~(\ref{eq:visc}-\ref{eq:NS}).  The shear rate profiles show the development
of a transient shear bands having two clearly distinct bands evolving towards a
homogeneous flow.  Comparing the velocity profiles, Fig~\ref{FIG3}(c), with the
carpobol gel experiments \cite{divoux} shows close similarities, even if the
model here exhibits no yield stress.

The origin of transient shear banding is a self-feeding mechanism, which can be
understood considering the Eqs.~(\ref{eq:simple}-\ref{eq:NS}).  When the fluid
starts to evolve at a homogeneous high volume fraction, the $\phi(r)$ decreases
fastest at the regions, where $\dot\gamma(r)$ has the largest value
(Eq.~(\ref{eq:simple})): close to the inner cylinder of the device.  The
accelerated decrease of $\mu(r)$ (Eq.~(\ref{eq:visc})) due to the faster
relaxation rate, further increases the $\dot\gamma(r)$ (Eq.~\ref{eq:NS}) at the
same location.  This accelerates the decrease of the $\phi(r)$
(Eq.~(\ref{eq:simple})) at the same position creating a self-amplifying
mechanism for the growth of transient shear banding.  Since the steady state
viscosity is constant, the $\phi(r)$ will decrease elsewhere in the device,
only with slower rate due to lower $\dot\gamma(r)$.  Finally a homogeneous
steady state profile is reached.  If the feedback either from $\phi(r)$ to
$\dot\gamma(r)$ (Eqs.  (\ref{eq:visc}) and (\ref{eq:NS})), or from
$\dot\gamma(r)$ to $\phi(r)$ (Eq.  (\ref{eq:simple})) is not strong enough, the
transient shear banding does not appear.  The intensity of the feedback is
adjusted by the exponent $k$ and the derivative of the Eq.~(\ref{eq:visc}) at
the corresponding $\phi(r)$.  If $\phi(r)$ at startup is below the steady state
one, such a feedback loop does not exist; the growth of the $\phi(r)$ is
fastest at the high shear rates, thus promoting the increase of the $\mu(r)$
and the decrease of the $\dot\gamma(r)$ at high shear rate regions.  This
reduces the shear rate differences in the gap.

\begin{figure}[htb]
\begin{center}
  \epsfig{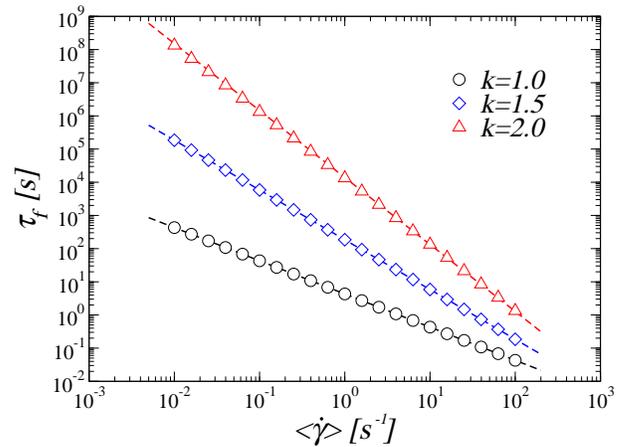}
\end{center}
\caption{\label{FIG4}  (Color online) The fluidization times against
$\langle \dot\gamma\rangle$.  
}
\end{figure}

It has been observed experimentally, that in simple yield stress fluids the
fluidization time decays with the global shear rate following a power-law. 
Similar behaviour is observed here (Fig.  \ref{FIG4}).
Here, the power-law decay follows $\tau_f \sim \langle \dot\gamma
\rangle^{-k}$, with $k$ fixed by/to the exponent $k$ in Eq.~(\ref{eq:simple}).  The
other parameters or even the form of the viscosity function
(Eq.~(\ref{eq:visc})) do not influence the power-law slope.  The rest of the
model parameters and the gap width simply change the vertical position of the
resulting power-laws.

The combination of Eq.~(\ref{eq:simple}) with $A_b=0$ and Eq.~(\ref{eq:NS}) that describes the
fluid in a Couette can be rewritten as
\begin{equation}
 \left(\frac{\dot\gamma_0}{\langle \dot\gamma \rangle}\right)^k \frac{d\phi(r,t)}{dt} = \left( A_s - B_s\phi(r,t) \right) \left[ F(r,t) \right]^k,
\end{equation}
where $\left[F(r,t)\right]^k$ represents the geometry effects, and in
particular is crucial for creating the transient shear bands.  This form
reveals a natural way to rescale the time: $t \langle \dot\gamma \rangle^k$. 
The measures dependent on the angular velocity can be rescaled with the global
shear rate.

\begin{figure}[htb]
\begin{center}
  \epsfig{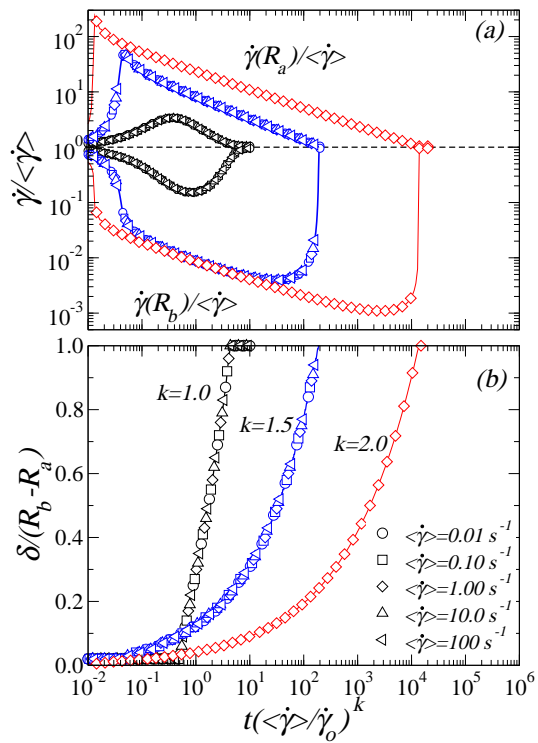}
\end{center}
\caption{\label{FIG5}  (Color online) The normalized shear rate evolution close
to the inner and outer cylinder (upper right panel), and the normalized
evolution of the shear band edge position in the gap.  }
\end{figure}

Other ways to illustrate transient shear banding, besides the velocity profiles,
are to plot the temporal evolution of the local shear rate at the gap edges,
and the band width \cite{divoux}.  These quantities are plotted with the same
rescaling of time and shear rate in Fig.~\ref{FIG5} (a) and (b), respectively. 
The band edge is estimated as the position, where $\left.  \frac{\partial^2
\dot\gamma(r,t)}{\partial r^2} \right|_{\delta} = 0$.  The Fig.~\ref{FIG5}
shows that there is a short induction period at small times, during which the
shear rate localizes near the rotor.  During the relaxation period, the shear
rate decreases towards the steady state value as the transient shear bands vanish.

Divoux {\it et al.} \cite{divoux_sm2011} demonstrated that the fluidization
time exponents in stress and shear rate controlled experiments relate to the
steady state power-law exponent.  Here, this equals unity.  As one would
expect, the fluidization for the stress controlled case follows a power-law
$t_f \sim \sigma^{-k}$ as does the shear rate controlled one $t_f \sim
\langle\dot\gamma \rangle^{-k}$.  A structural kinetics model producing
Herschel-Bulkley flow curve with a control of the exponent can be speculated to
change also the the fluidization exponents towards the experimental ones
\cite{divoux_sm2011}.  To the best of our knowledge, such a model does not
exist.  Studying the properties of more advanced structural kinetics equations
on the fluidization is therefore left to future studies.

{\it Conclusions}. --- We have studied transient shear banding, or fluidization of a simple time
dependent fluid.  We constructed a minimal model which lacks further
complications such as the (visco)elasticity of the structure, a
physically-motivated yield stress, the normal stresses suggested very recently
\cite{cheddadi} to play a role in steady state shear banding~\cite{moller}, or
elastic stress overshoots \cite{fielding}. Our analysis indicates
that such conditions are not required for transient shear banding. On the
contrary, transient shear banding should be a general feature of complex fluids
like colloidal suspensions and microgels.

The transient shear bands during the fluidization was found to originate from
the initial shear inhomogeneity here arising from the Couette geometry, which
is amplified by the positive feedback coming from the shear dependent
relaxation rates of the fluid.  This mechanism should be present in all similar
scenarios of fluidization: studying the couplings built into the model implies,
that the same effect should occur in all practical measuring geometries for
shear thinning complex fluids.  This is since even the smallest stress gradient
is enough to trigger the transient shear bands due to the shear dependent response of the fluid. 
Experimental observations of stress signatures associated to transient shear
banding found in a cone-and-plate geometry, where the stress gradient is
extremely small, also support this finding \cite{divoux_sm2012}.
Furthermore, in Eq.~\ref{eq:NS} we neglected the inertial terms in the Navier-Stokes
equation, which could play a role in the start-up flow in such situations
(referring to cone-and-plate).  The self-amplifying mechanism between the flow
and the fluid structure would be the same, but the origin of the shear
inhomogeneity would come from the inertial terms rather than the flow geometry,
which was the case in the circular Couette studied here.

The kinetic exponent related to the dependence of $\phi$ on $\dot\gamma$, is
connected to the fluidization exponent, in both the stress and shear controlled
cases.  This indicates that studying the fluidization experimentally gives
detailed information of the timescales of the internal relaxation processes in
time-dependent fluids, which could be utilized to build proper structural
models based on the experimental fluidization data. 
Such models could be established for instance around rheological models describing
the volume fraction using Population Balances, which take into account the particle 
size distribution and concentration as rheological parameters \cite{heath}. Presently, work is
devoded along these lines of research.  

\section{Acknowledgements}
We are grateful to S. Manneville and T. Divoux for fruitful discussions. 
This work has been supported by the Effnet program in  the Finnish Forest
Cluster Ltd, and EU Framework 7 program SUNPAP. Also, the support
from the Academy of Finland through the COMP center of excellence
and the project number 140268 and within the framework of the
International Doctoral Programme in Bioproducts Technology (PaPSaT)
are acknowledged.

\section{References}
\bibliographystyle{apsrev}

\end{document}